\documentstyle[aps]{revtex}

\begin{document}

\begin{flushright}
e-print JLAB-THY-97-23 \\
June 1997 \\
%hep-ph/
\end{flushright}

\vspace{2cm}

\begin{center}
{\Large \bf
Deeply Virtual Compton Scattering: Facing Nonforward 
Distributions}
\end{center}
\begin{center}
{\sc Anatoly V. Radyushkin\footnote{Also at Laboratory of Theoretical Physics,
JINR, Dubna, Russia}} \\ 
{\em Physics Department, Old Dominion University,
 Norfolk, VA 23529, USA}  \\ {\em
 Jefferson Lab, Newport News,VA 23606, USA}

\end{center}
\vspace{2cm}

\begin{abstract}
Applications of perturbative QCD to
deeply virtual Compton scattering
process require a generalization of 
usual parton distributions for the case when
long-distance information
is accumulated in  nonforward matrix
elements of quark and gluon   operators.
We discuss
two types of functions 
parametrizing such matrix elements:
double distributions $F(x,y;t)$ and 
nonforward distribution functions
${\cal F}_{\zeta}(X;t)$ and also their relation to 
usual parton densities $f_a(x)$.
\end{abstract}

\section*{Proton spin and DVCS}

Recently, X. Ji \cite{ji1} suggested to use deeply virtual Compton scattering 
(DVCS) to get information about the total 
quark and gluon contributions to the spin of the proton.
Note, that  the angular momentum operator $ J^{\mu\nu}$ can be written 
in terms of the symmetric  energy-momentum tensor 
$T^{\alpha\beta}$:
\begin{equation}
       J^{\mu\nu} = \int d^3 \vec{x}~ M^{0\mu\nu}(\vec{x}) \ \ ; \  \  
      M^{\alpha\mu\nu} = T^{\alpha\mu} x^\nu
         - T^{\alpha\nu}x^\mu \ ,   
\label{1} 
\end{equation}
which in QCD can be represented as a sum of quark and gluon parts
\begin{equation}
      T^{\alpha\beta}  =
          T_q^{\alpha\beta} + T_g^{\alpha\beta} 
      = {1\over 4}\bar \psi \gamma^{\{\alpha} 
        i\stackrel{\leftrightarrow}{D^{\beta \} }} \psi
         + \left({1\over 4}g^{\alpha\beta}F^2 -F^{\alpha\mu}
          F^\beta_{~\mu}\right) \  .  
\label{2}
\end{equation}
Hence, the quark $J_q$ and gluon $J_g$ contributions to the proton 
spin can be obtained from the form factors $A_{q,g}(t), B_{q,g}(t)$ of the 
energy-momentum tensor at zero momentum transfer $t=0$:
\begin{equation}
J_{q,g} = \frac12 \left [ A_{q,g}(0)+ B_{q,g}(0) \right ] \ \ ; 
 \  \ J_{q}+J_{g} = \frac12 \ . 
\end{equation}
The functions  $A_{q,g}(t), B_{q,g}(t)$ are defined by \cite{ji1}
\begin{eqnarray} 
\langle p' | T_{q,g}^{\mu\nu} | p \rangle = \bar u (p') \left [
A_{q,g}(t) \gamma^{\{ \mu} P^{\nu \}} +  B_{q,g}(t) 
 P^{\{\mu } \sigma^{\nu \} \alpha} \frac{r_{\alpha}}{2M}\right. \nonumber \\
\left.  + C_{q,g}(t) (r^{\mu}r^{\nu} -g^{\mu \nu}t) + 
D_{q,g}(t) M g^{\mu \nu} \right ]u(p)
\end{eqnarray} 
where $r \equiv p'-p, t \equiv r^2$. To get $J_{q,g}$, one should know
both the non-spin-flip quantities   $A_{q,g}(0)$ related to total
hadron momentum carried by quarks or gluons and spin-flip amplitudes 
invisible in deep-inelastic cross sections corresponding
to exactly forward $r=0$ virtual Compton amplitude. 
However, information about $B_{q,g}(t)$ and $A_{q,g}(t)$,
is contained in the DVCS amplitude  \cite{ji1}. 
Anyway, whether the extraction of $B_{q,g}(0)$ 
 is feasible or not,   the 
studies of DVCS, an elastic process
exhibiting, in the Bjorken limit, a scaling behavior  similar to that of DIS,
may be interesting on its own grounds \cite{compton,ji2}
(earlier discussions of nonforward Compton-like   
amplitudes $\gamma^* p \to \gamma^* p' $ 
with a  virtual photon or $Z^0$ in the
final state  were given    in refs. \cite{barloe,glr,drm}).  

\section*{ Parton picture: double distributions}

The  kinematics  of the process $\gamma^* p \to \gamma  p'$ is 
specified  
by the initial nucleon momentum $p$, virtual photon momentum $q$
and    momentum  transfer $r=p-p'$. 
We are interested in  the situation, when the invariant momentum transfer 
$t \equiv r^2$ and  $p^2\equiv m_p^2$ 
are much smaller than  
the  virtuality $q^2  \equiv -Q^2 $ of the initial 
photon and the energy invariant $p \cdot q $,
with the Bjorken ratio $Q^2/2(p \cdot q) \equiv x_{Bj}$ fixed.
It is helpful  to consider first a (formal) limit 
$p^2=0$ and $t = 0$ and take $p$ as a basic light-cone 
momentum in the Sudakov decomposition.
 Since 
 the final photon momentum $q'$ is  light-like  $q'^2 =0$, 
it is natural to use $q'$ 
as another  basic light-cone  4-vector. Then $q= q' - x_{Bj} p$.
Furthermore,  in this limit, the  requirement
$p'^2 \equiv (p+r)^2=p^2$  reduces to the condition 
$p\cdot r = 0$  which can be satisfied only
if the  two lightlike momenta
$p$ and $r$ are proportional to each other: $r= \zeta  p$,
where $\zeta$ coincides with 
$ x_{Bj}$ 
which satisfies  $0 \leq x_{Bj}  \leq 1$.

The leading contribution in the  large-$Q^2$, fixed-$x_{Bj} $,
small-$t$  limit
is given by DIS-type handbag
diagrams in which  the long-distance 
dynamics is described by  matrix elements 
$
\langle p -r  \,  | \,   \bar \psi_a(0) \gamma_{\mu} 
E(0,z;A) \psi_a(z) \, | \,p  \rangle $ 
and  \mbox{$
\langle p -r  \, | \, \bar \psi_a(0) \gamma_5 \gamma_{\mu} 
E(0,z;A) \psi_a(z) \, | \,p \rangle , 
$}
where  $E(0,z;A)$ is the usual
$P$-exponential of the gluonic $A$-field  
along the straight line  connecting 0 and $z$.
Though the momenta
$p$ and $r$ are proportional to each other $r = \zeta  p$,
to construct an adequate QCD parton picture,
one should make a clear distinction between them.
The basic  reason 
is  that $p$ and $r$  specify 
the momentum flow in two different   channels.
For $r=0$, the net momentum flows only in the $s$-channel 
and the total momentum entering into 
the  composite operator vertex is zero. 
In this case, 
the matrix element coincides with  the 
standard   distribution function.
The  partons  entering the composite vertex 
then carry  the   fractions $x_i$ 
of the initial proton momentum ($-1 < x_i <1$). 
When  $x $ is  negative, the parton is  interpreted
as belonging to the final state and  $x_i$ is redefined
to  secure that  
the integral always  runs over  the segment  $0\leq x\leq 1$.
In this parton picture, the spectators take the 
remaining momentum  $(1-x)p$.
On the other hand, if  the 
total momentum flowing through the 
composite vertex is $r$, 
the matrix element has the structure
of the distribution amplitude in which 
the momentum  $r$  
splits into the fractions $yr$  and 
$(1-y)r \equiv \bar y r$ carried by the 
quark fields  attached to  that vertex. 
In a combined situation, when both $p$ and $r$
are nonzero,  the initial quark  has the momentum 
$xp +y r$, while the final one  carries the momentum 
$xp - \bar y r$.
In more formal terms, this corresponds  
to the following parameterization 
of the light-cone matrix elements 
\begin{eqnarray} 
&& \langle p-r\, | \, \bar \psi_a(0) \hat z 
E(0,z;A)  \psi_a(z) \, | \, p \rangle |_{z^2=0} 
 =  \bar u(p-r)  \hat z 
u(p)  \int_0^1   \int_0^1  \, \theta( x+y \leq 1) \nonumber \\
&& \hspace{2cm}  \left ( e^{-ix(pz)-iy(r z)}F_a(x,y;t) 
  -  e^{ix(pz)-i\bar y(r z)}F_{\bar a}(x,y;t)
\right )
 \,  dy \, dx , \label{eq:vec} \\
&& \langle p-r\, | \, \bar \psi_a(0) \gamma_5 \hat z 
E(0,z;A)  \psi_a(z) \, | \, p \rangle |_{z^2=0}  
 =  \bar u(p-r)  \gamma_5 \hat z 
u(p)  \int_0^1   \int_0^1 \theta( x+y \leq 1) \nonumber \\ 
&& \hspace{2cm}  \left ( e^{-ix(pz)-iy(r z)}G_a(x,y;t) 
 +   e^{ix(pz)-i\bar y(r z)}
G_{\bar a}(x,y;t) \right ) \, 
 dy \, dx ,
\label{eq:ax}
\end{eqnarray} 
where $\hat z \equiv \gamma_{\mu} z^{\mu}$. 
Though  we arrived at the matrix elements  (\ref{eq:vec}), (\ref{eq:ax})
in the context of the scaling limit of the DVCS amplitude,
they accumulate a  process-independent information.
The coefficient of proportionality between 
$(pz)$ and $(rz)$ is  just a parameter  characterizing
``skewedness'' of the matrix elements.
The fact that, in our case, $\zeta$ coincides with 
the Bjorken variable is  specific  for 
the DVCS amplitude.
An important  feature  implied by
the representation (\ref{eq:vec}),(\ref{eq:ax}) is the absence 
of the $\zeta$-dependence in 
the double distributions $F_a(x,y;t)$ and $G_a(x,y;t)$.
This property and 
spectral constraints $x \geq 0$,
$y \geq 0$, $x+y \leq 1$ 
 hold for  any Feynman diagram \cite{evol}.
As a result,  both the initial active quark 
and the spectators  carry  positive 
fractions of the light-cone ``plus'' momentum $p$:
$x+\zeta y$ for the active quark and 
$(1-x-y) +(1-\zeta )y$ 
for   the spectators. 
However,  the fraction of the initial momentum  $p$
carried by the ``returning''
quark is given by 
$x - \bar y \zeta $ and it may take  both 
positive and negative values.

Taking the  limit $r =0$ gives  the matrix
element defining the parton distribution functions 
$f_{a,\bar a}(x)$, 
$\Delta f_{a, \bar a}(x)$.
This  observation results   in the following reduction formulas
 for  the double distributions $F(x,y;t), G(x,y;t)$:
\begin{equation}
\int_0^{1-x} \, F_a(x,y;t=0)\, dy=  f_a(x) \  \ ,  \  \ 
\int_0^{1-x} \, G_a(x,y;t=0)\, dy=  \Delta f_a(x).
\label{eq:redf}
\end{equation}

The parameterization for the matrix elements
given above, results in  a parton  representation
for the handbag contributions to the DVCS amplitude:
\begin{eqnarray} 
& & T^{\mu \nu} (p,q,q') =  \frac{ 1}{2\, (p q')} \,
\sum_a 
e_a^2\, \left [
\left (-g^{\mu \nu} + \frac1{p \cdot q' } 
(p^{\mu}q'^{\nu} +p^{\nu}q'^{\mu}) \right ) 
 \biggl \{  \bar u(p') \hat q'  u(p) T_F^a(\zeta )   \right. 
\nonumber \\ & & \left. 
+  \frac1{2M} \bar u(p') 
(\hat q' \hat r - \hat r \hat q' )u(p)  T_K^a(\zeta ) + \{ a \to \bar a \}
\biggr \} \right.
\label{122}  \\ & & \left.
+ i \epsilon^{\mu \nu \alpha \beta} \frac{p_{\alpha} q'_{\beta}}{(pq') }
\,  
\left \{ \bar u(p') \gamma_5 
\hat q'  u(p) 
\,  T_G^a(\zeta ) 
+  \frac{(q'r)}{2M} \bar u(p') \gamma_5 
  u(p)  T_P^a(\zeta ) + \{ a \to \bar a \} \right \}  \right ] \nonumber
 \end{eqnarray}
where 
$T^a(\zeta,t )$ are the 
 functions 
depending on  the scaling variable $\zeta $:
\begin{eqnarray} 
T_F^a(\zeta,t ) = -  \int_0^{1}  dx \, \int_0^{1-x} \left
 ( \frac1{x-\zeta \bar y+i\epsilon}
+ \frac1{x+\zeta  y} \right ) F_a(x,y;t) \,  dy , 
\label{eq:tv} 
\end{eqnarray} 
{\it etc.} The terms  containing $1/(x-\zeta \bar y+i\epsilon)$ 
generate the imaginary part:
\begin{eqnarray} 
 \frac1{\pi}\, {\rm Im} \, T_F^a(\zeta ,t) =  
 \int_0^1
F_a( \bar y \zeta ,  y ;t) \, d  y , 
\label{eq:imtv}
\end{eqnarray} 
with a similar expression for $ {\rm Im} \, T_A^a(\zeta ,t)$.
The  relation between 
${\rm Im} \, T(\zeta,t )$ and the double distributions $F_a(x,y;t)$
is not as direct as in the case of forward virtual 
Compton amplitude, the imaginary part of which 
is just  given by parton densities $f_a(\zeta)$. 
Note, that the $y$-integral in eq.(\ref{eq:imtv}) 
 is different  from that in the reduction formula 
(\ref{eq:redf}), $i.e.,$ though 
\begin{equation}
\Phi_a(\zeta,t) \equiv \int_0^1
F_a( \bar y \zeta ,  y ;t) \, d  y 
\label{Phi}
\end{equation}
is a function of the Bjorken  variable $\zeta$, it does 
not coincide with $f_a(\zeta)$.
To get the real part of the $1/(x-\zeta \bar y+i\epsilon)$ terms,
one should use the principal value prescription,
$i.e.,$ ${\rm Re} \, T(\zeta ,t)$ is related to 
$F_a(x,y;t)$ through two integrations.  

\section*{Nonforward Distributions} 

Since $(rz) = \zeta (pz)$,  the variable $y$ appears 
in eqs.(\ref{eq:vec}),(\ref{eq:ax}) only in the 
$x+y\zeta \equiv X$ 
combination,
where $X$  can be treated as    the {\it total}  fraction 
of the initial hadron momentum $p$ carried by the active  quark.
Since $\zeta \leq 1$ and  $x+y \leq 1$, 
the variable $X$ satisfies a  natural
constraint $0\leq X \leq 1$.
Integrating the  double   distribution $F(X-y \zeta,y)$ 
over $y$ gives  the {\it nonforward parton distribution }  \cite{gluon,evol}
\begin{equation}
{\cal F}_{\zeta} (X;t) = 
 \int_0^{\min \{ X/\zeta, \bar X / \bar \zeta \} } F(X-y \zeta,y;t) \, dy
\label{25} \end{equation}
where $\bar \zeta \equiv 1- \zeta$.
The  basic distinction  between 
the  double  distribution $F(x,y;t)$ 
and the  nonforward  distribution 
${\cal F}_{\zeta} (X;t)$ is that  
the latter   explicitly depends on the skewedness  parameter
$\zeta$:
 one  
deals  now  with a family of 
nonforward distributions  ${\cal F}_{\zeta} (X,t)$ 
whose shape changes when $\zeta$ is changed.

The fraction $X- \zeta \equiv X' $ of the
original hadron momentum $p$  carried by the ``returning''
parton differs from $X$ by $\zeta$: $X-X' = \zeta $.
Since $X$ changes from $0$ to $1$ and $\zeta \neq 0, 1$, the fraction 
$X'$ can be either positive  or negative, $i.e.,$  
 the asymmetric 
distribution function has two components
corresponding to the regions $1 \geq X \geq \zeta$ and 
 $0 \leq X \leq \zeta$.
In the region $X > \zeta$, 
the function ${\cal F}_{\zeta} (X)$ 
can be treated as  a generalization of the 
usual  distribution function $f(x)$
for the asymmetric case when the final hadron momentum $p'$ 
differs by $\zeta p$ from the initial momentum $p$.  
In the region  $X < \zeta$ 
 the ``returning''  parton   has
a negative fraction $(X- \zeta)$ of the light-cone momentum $p$.
Hence, it is more appropriate to  treat it  as a parton 
going out of the hadron and 
propagating  along  with the original parton.
Writing $X$ as $X = Y \zeta$, we see that 
both   partons  carry now 
positive fractions $Y \zeta p \equiv Y r$ and
  ${\bar  Y} r  \equiv (1-Y)\, r $ of 
the momentum transfer $r$. Thus, the 
nonforward  distribution 
in the region $X= Y \zeta < \zeta$
looks like a distribution amplitude 
$\Psi_{\zeta}(Y;t)$ for a  $\bar q  q$-state 
with the  total momentum $r= \zeta p$: 
\begin{equation}
\Psi_{\zeta}(Y;t) =  \int_0^Y F((Y-y) \zeta , y ;t) \, dy . 
\label{28} \end{equation}

In terms of ${\cal F}_{\zeta}(X)$, 
  the $F$-part of the virtual Compton amplitude 
 is
\begin{eqnarray} 
T_F^a(\zeta ) =
- \int_0^{1} 
\left  [ \frac1{X-\zeta +i\epsilon}
+ \frac1{X- i \epsilon} \right ]  \left ( {\cal F}^a_{\zeta}(X;t)
+ {\cal F}^{\bar a}_{\zeta}(X;t) \right ) \, dX \,  . 
\label{123}  \end{eqnarray} 
It can be shown that 
 ${\cal F}_{\zeta}^{a, \bar a}(X)$ linearly vanish as $X \to 0$ \cite{evol}. As 
a result, the imaginary part  is generated 
by the $1/(X-\zeta + i \epsilon)$ singularity 
\begin{eqnarray} 
\frac1{\pi}\, {\rm Im} \, T_F^a(\zeta ) = {\cal F}^a_{\zeta}(\zeta;t) \, + \, 
{\cal F}^{\bar a}_{\zeta}(\zeta;t) 
\label{125} \end{eqnarray} 
Hence, the integral $\Phi(\zeta,t)$ in Eq.(\ref{Phi})  is equal to
${\cal F}_{\zeta}(\zeta;t)$, $i.e.,$
to the nonforward  distribution  ${\cal F}_{\zeta}(X;t)$
taken at the point $X = \zeta$. The parameter $\zeta$ 
is present  in ${\cal F}_{\zeta}(\zeta;t)$ twice: first as 
the parameter specifying the skewedness 
of the matrix element and then as 
the momentum fraction at which  the imaginary part appears.
As one may expect, it appears for $X = x_{Bj} =\zeta$,
just like in the forward case. 
Note, however, that  the momentum $(X- \zeta) p$ of 
the ``returning'' parton vanishes when $X  = \zeta$:
 the imaginary part 
appears in a highly asymmetric  
configuration in which the
fraction
of the original hadron momentum carried
 by the second parton 
vanishes.
Hence,  
${\cal F}_{\zeta}(\zeta)$ in general 
 differs from the function 
$ f(\zeta)$.  
Experimentally, the imaginary part of the DVCS 
amplitude can be extracted by measuring the 
single-spin asymmetry \cite{ji2}.

{\it Acknowledgement} This work was supported by the US Department of Energy
 under contract 
DE-AC05-84ER40150.

\end{document}